\pdfoutput=1
\documentclass[prd,twocolumn,nofootinbib,reprint]{revtex4}
\usepackage{amsmath,amssymb,amsthm,bm,braket,ascmac,url,hyperref,graphicx,color,comment} 
\newcommand{\be}{\begin{eqnarray}}
\newcommand{\ee}{\end{eqnarray}}
\newcommand{\R}{\mathcal{R}}
\newcommand{\pbar}{\bar p}
\newcommand{\pb}{\bar p}
\newcommand{\ep}{e^+}
\newcommand{\ixs}{E\frac{d^3\sigma}{dp^3}}

\begin{document}
\title{$\boldsymbol{\ep}$ and $\boldsymbol{\pb}$ production in $\boldsymbol{pp}$ collisions and the cosmic-ray $\boldsymbol{e^+/\pbar}$ flux ratio
}
\author{Kfir Blum$^{1,2}$, Ryosuke Sato$^1$ and Masahiro Takimoto$^{1,3}$}
\affiliation{\vspace{2mm} $^1$Department of Particle Physics and Astrophysics, Weizmann Institute of Science, Rehovot 7610001, Israel \\
$^2$CERN, Theoretical Physics Department, Geneva, Switzerland\\
$^3$Institute of Particle and Nuclear Studies, High Energy Accelerator Research Organization (KEK),\\
Tsukuba 305-0801, Japan}

\date{\today}

\begin{abstract}
\vspace{1mm}
Secondary astrophysical production of $e^+$ and $\pbar$ cosmic rays is considered.
Inclusive $\pi$, $K$, and $\pbar$ production cross sections in $pp$ collisions at large $\sqrt{s}$ are parametrised using recent experimental data at LHC energies.
The astrophysical production rate ratio $Q_{e^+}/Q_{\pbar}$ is calculated for an input cosmic ray proton flux consistent with local measurements. At $10<E<100$~GeV the cosmic ray flux ratio $J_{e^+}/J_{\pbar}$ measured by AMS02 falls below the production rate ratio by about 50\%, while at high energy $E>100$~GeV the measured flux ratio coincides with the production rate ratio of the secondary source.
\end{abstract}
\maketitle

%#######################
\section{Introduction}\label{sec:intro}
Cosmic ray (CR) antimatter is a potential probe of exotic high energy astrophysical phenomena and a unique diagnostic of CR propagation. 
Over the last decade, precise measurements of the flux of CR $e^+$ and $\bar p$ extending to ever higher energies were reported by the PAMELA and AMS02 experiments \cite{Accardo:2014lma, Aguilar:2014mma,Aguilar:2016kjl}.
The interpretation of these measurements motivates refined theoretical consideration of astrophysical $e^+$ and $\bar p$, produced as secondaries in the collision of primary CRs, notably protons, with interstellar matter (ISM), notably hydrogen. 
Our goal in the current paper is to improve on previous calculations of the inclusive production cross section of secondaries in $pp$ collisions using recent accelerator data.

The main effect we wish to capture is the violation of radial scaling at $\sqrt{s}>50$~GeV. As shown in Refs.~\cite{diMauro:2014zea, Winkler:2017xor, Donato:2017ywo}, this effect leads to about a factor of two increase in the astrophysical $\pbar$ source at $\pbar$ energy above a few TeV. Here we evaluate the analogous effect in the CR $e^+$ source by analysing meson production at LHC energies. Earlier  $e^+$ calculations were either based on too low $\sqrt{s}$ data to see the effect~\cite{Moskalenko:1997gh, Delahaye:2008ua, Kappl:2016qug} or relied on Monte-Carlo tools without direct verification in the kinematical regime relevant for astrophysics~\cite{Kamae:2006bf}.

We aim to achieve $\sim$10\% accuracy for the astrophysical $e^+$ source at $e^+$ energy ranging from a few GeV up to multi-TeV; this accuracy goal is to be compared with the main radial scaling violation effect that is, again, about a factor of two at $E\sim10$~TeV. As a check against earlier work, we also calculate the $\pbar$ source to similar accuracy. 

In section \ref{sec:radialscaling} we analyse the cross sections at large $\sqrt{s}$,  using results from the NA49, PHENIX, ALICE, and CMS experiments.
In section \ref{sec:general_bound} we use these results to calculate the production rate ratio $Q_{\ep}/Q_{\pb}$ for secondary $\ep$ and $\pb$ produced by a spectrum of high energy protons scattering on a proton target. 
%The production rate ratio provides an upper bound on the secondary astrophysical $\ep/\pb$ flux ratio. 
We show that $Q_{\ep}/Q_{\pb}$ is insensitive w.r.t. uncertainties in the primary proton spectrum. At $10<E<100$~GeV the $\ep/\pb$ flux ratio measured by AMS02 falls bellow the production rate ratio by about 50\%, while at high energy $E>100$~GeV the measured flux ratio coincides with the production rate ratio of the secondary source. 
In App.~\ref{sec:kaon} we discuss the contribution to secondary $\ep$ from $K_L^0$ decay, which was missing in previous calculations. In App.~\ref{app:ant} we analyse the hyperon contribution to inclusive $\pb$ production. In App.~\ref{app:secpb} we reproduce the secondary cosmic ray $\pb$ flux predicted by using mean traversed target column density as deduced from cosmic ray nuclei data.

\section{Data analysis}\label{sec:radialscaling}
Our baseline fitting formulae for inclusive hadron production in $pp$ collisions are taken from Ref.~\cite{Tan:1984ha} (Tan\&Ng), which was based on $\sqrt{s}\leq53$~GeV data and to which we provide corrections using the following new information:

{\bf i.} Tan\&Ng's formulae rely on radial scaling \cite{Feynman:1969ej, Kinoshita:1971uu, Kinoshita:1973gd},
\begin{align}
\ixs(x_R,p_t,\sqrt{s})
\to_{\sqrt{s}\to\infty}
\ixs(x_R,p_t)
\end{align}
where $x_R=E^*/E^*_{\rm max}$, $E^*$ is the final state hadron energy in the centre of mass (CM) frame 
and $E^*_{\rm max}$ is the maximum attainable $E^*$. 
Recent accelerator data show violation of radial scaling in $pp$ collisions at  $\sqrt{s}\gtrsim50$~GeV~\cite{Aamodt:2011zj,Abelev:2014laa,Adam:2015qaa,Chatrchyan:2012qb,Adare:2011vy}. 
The $pp\to\bar p$ cross section increases at high energy~\cite{diMauro:2014zea, Winkler:2017xor, Donato:2017ywo} as compared to \cite{Tan:1984ha} and other early parametrisations. We will assess the analogous effect in meson production and the resulting $e^+$ yield.

{\bf ii.} In addition to the high energy end, unprecedented detailed measurements of the production cross section $\pi^+$, $K^\pm$ and $\bar{p}$~\cite{Alt:2005zq,Anticic:2010yg,Anticic:2009wd} at $\sqrt{s}=17.2~$GeV were reported by the NA49 experiment. This value of $\sqrt{s}$ is particularly relevant for $E\sim10-100$~GeV final state $\pb$ and $\ep$~\cite{Blum:2017qnn}. We incorporate this data in our formulae for hadronic cross sections.

\subsection{$p_t$-weighted cross sections and important kinematical region}
Faced with an extensive data set~\cite{Aamodt:2011zj,Abelev:2014laa,Adam:2015qaa,Chatrchyan:2012qb,Adare:2011vy,Alt:2005zq,Anticic:2010yg,Anticic:2009wd}, it is instructive to bracket the final state phase space that is most relevant for secondary CR production. 
In the fixed-target set up of high energy CR scattering on ambient ISM, the key quantity is the conversion cross section 
from incoming CR proton with ISM frame energy $E_p$ to outgoing secondary particle with ISM frame energy $E$,
\begin{align}
         \frac{d\sigma(E_p,E)}{dE}=2\pi\int_0^{\pi}d\theta p_t
         \left(\ixs\right)(x_R,p_t,\sqrt{s}) ,
\end{align}
where $\theta$ denotes the angle between the incoming proton and outgoing secondary in the ISM frame.
The Lorentz-invariant differential cross section $\ixs$ decreases sharply with increasing $p_t$, with the $p_t$-weighted cross section $p_t\left(\ixs\right)$ peaking around average $\langle p_t\rangle\sim0.2-0.4$~GeV. 
For $E\gg  \langle p_t\rangle,\,m$, where $m$ is the mass of the final state hadron of interest, we can simplify the integral as
\begin{align}
\label{eq:avp}
       \int_0^{\pi}d\theta &p_t
         \left(\ixs\right)(x_R,p_t,\sqrt{s}) \nonumber \\ &\simeq
\frac{1}{p}\int_0^{\infty}dp_t~p_t
         \left(\ixs\right)(x_R|_{p_t=0},p_t,\sqrt{s}) ,
\end{align}
where $x_R|_{p_t=0}$ is computed at $p_t=0$ and only depends on $E$ and $E_p$.
%Corrections to this approximation are of the order of $\langle p_t\rangle^2/Em_p$. 
This exercise shows that,
in the high energy regime,
 the $p_t$-weighted mean cross section with fixed $x_R$
is the most important quantity for secondary CR production, allowing one to average over the detailed $p_t$ dependence reported by the experiments.

Next, we consider the relevant range of $x_R$.
Consider as a representative example the cross section parametrization~\cite{Winkler:2017xor}:
\begin{align}\label{eq:xr}
         E\frac{d^3\sigma}{dp^3}(x_R,p_t)=f_0e^{-\frac{p_t}{\langle p_t\rangle}}
         (1-x_R)^n.
\end{align}
Typical parameters are $\langle p_t\rangle\simeq 0.2$--$0.4~$GeV and $n\simeq 5$--$7$.
In the limit where $m^2/E^2,~p_t^2/E^2 \ll 2m_p/E_p$,
the astrophysical source term $Q(E)$ can be written as
\begin{align}
Q(E)\propto &\int_E^\infty d E_p J_p(E_p)  \frac{d\sigma(E_p,E)}{dE}\nonumber \\
         &\simeq2\pi f_0\langle p_t\rangle^2 J_p(E)\int_0^1 dx_R ~x_R^{\gamma-2}(1-x_R)^n,
         %\nonumber \\
         %&=2\pi f_0\langle p_t\rangle^2 n_p(E)B(\gamma-1,n+1),
\end{align}
where $J_p$ denotes the CR proton flux and we assumed $J_p\propto E_p^{-\gamma}$.
With $n\simeq 5$--$7$ and $\gamma \simeq 2.7$, the $x_R$ integrand selects the range 
$\sim 0.1$--$0.4$.

To summarise, we are most interested in the cross section for secondary product energy in the range $E\gtrsim 10~$GeV. In this range, the relevant information is contained in the $p_t$-weighted mean invariant cross section at fixed $x_R$, where furthermore the relevant range of $x_R$ is $\sim$0.1--0.4.

\subsection{Hadron production cross section}
In this section we discuss the hadronic cross section in the light of recent collider experiments.
We take the cross section fits by Tan\&Ng as baseline, and derive corrections to these formula.

A comment is in order regarding the intermediate hyperon contribution to $\bar p$.
In $pp$ collisions, $\bar p$ are generated promptly or by the decay of (relatively) long-lived hyperons, notably $\bar\Lambda$ and $\bar\Sigma^\pm$.
The Tan\&Ng $\pb$ fit includes the hyperon contributions.
On the other hand, recent experiments such as NA49 report the prompt antiproton cross section in which 
the contribution of intermediate hyperon states is removed.
Thus, when comparing experimental $\pb$ cross section data and fits 
we need to specify whether the hyperon contribution is subtracted or not.

For the purpose of astrophysical calculations, of course, 
our eventual concern is the total $\bar p$ cross section including the hyperon contributions. 
In this section, however, we find it convenient to concentrate first on the prompt $\pb$ production cross section, deferring an analysis of the hyperon contribution to App.~\ref{app:ant}.

\subsubsection{NA49 experiment}
\begin{figure}
\centering
\includegraphics[width=\hsize]{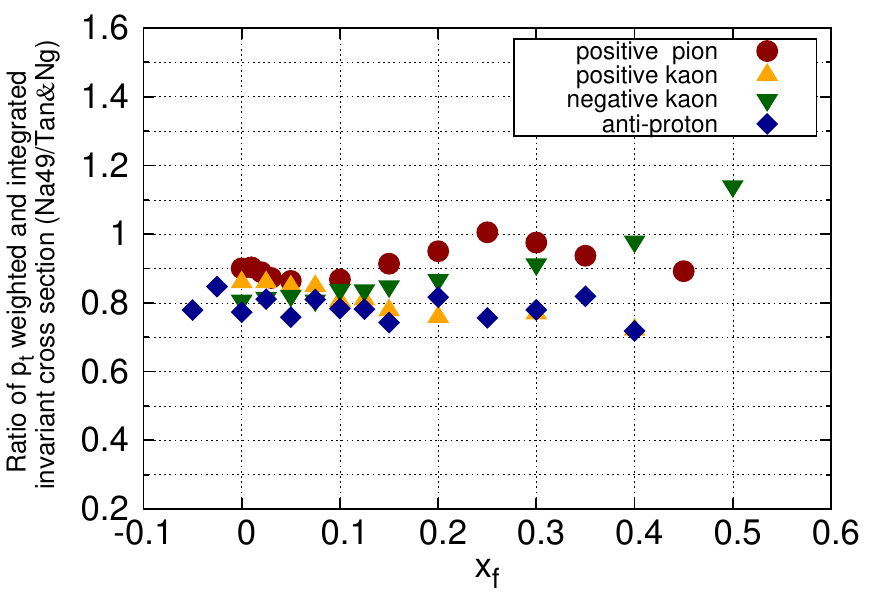}
\caption{
$p_t$-weighted cross section for $\pi^+$, $K^\pm$ and $\bar{p}$, presented as ratio
between NA49 data and Tan\&Ng~\cite{Tan:1984ha} inclusive cross section formulae, in given $x_F$ bins. Note that the Tan\&Ng formulae include contributions from unstable intermediate states, such as the hyperon contribution $\bar\Lambda\to\pbar\pi^+$ to $\pbar$ production, that is subtracted in the NA49 $\pbar$ data. We explain how to correct for this effect in the text.
}\label{fig:z49}
\end{figure}

The NA49 experiment reported measurements in a wide kinematic regime.
Fig.~\ref{fig:z49} shows measurements of the $p_t$-weighted cross section, presented as ratio between NA49 data and the Tan\&Ng's formulae in given $x_F$ bins\footnote{NA49 data are provided in terms of the Feynman parameter $x_F=2p_L^*/\sqrt{s}$ (where $p_L^*$ is the hadron longitudinal momentum in the CM frame) instead of $x_R$, so we consider the $p_t$-weighted cross section at fixed $x_F$.}. We use data from 
~\cite{Alt:2005zq},~\cite{Anticic:2010yg} and~\cite{Anticic:2009wd} for $\pi^+$, $K^\pm$ and $\bar{p}$ respectively. 
For each point, statistical and systematic errors are both at the level of 10\%.

As we discussed, the most relevant kinematical region to determine CR flux is $x_F = 0.1$--$0.4$.
In this region, Fig.~\ref{fig:z49} shows that
apart from an overall factor the fitting functions of Tan\&Ng are consistent with the NA49 results for all final states with the possible exception of $K^-$ (the latter being quantitatively irrelevant for the secondary $e^+$ calculation).
Motivated by this result, we introduce a scaling factor $\xi_H(\sqrt s)$ for each hadron $H=\pi^+,K^\pm,\bar p$,
and parametrize the cross section as
\begin{align}
E\frac{d^3 \sigma_H}{dp^3}
=
E\frac{d^3 \sigma_H}{dp^3} \biggr|_{\rm Tan\&Ng}%_{\rm Ref.~\cite{Tan:1984ha}}
\times \xi_H(\sqrt{s}).
\end{align}
We take $\xi_{\pi^+} = \xi_{K^\pm} = 0.9$ and $\xi_{\bar p} = 0.8$ at $\sqrt s = 17.2~{\rm GeV}$.

Note that the prompt $\pb$ cross section from NA49 is off by $\sim20\%$ from the inclusive Tan\&Ng fit: this is not a discrepancy, but is mainly due to the 
 hyperon contribution present in the Tan\&Ng fit while being subtracted from NA49 data. Accounting for this correction we find, in fact, that the inclusive Tan\&Ng fit is in good agreement with that deduced from NA49 data.

\subsubsection{High energy experiments}
\begin{figure}
\centering
\includegraphics[width=\hsize]{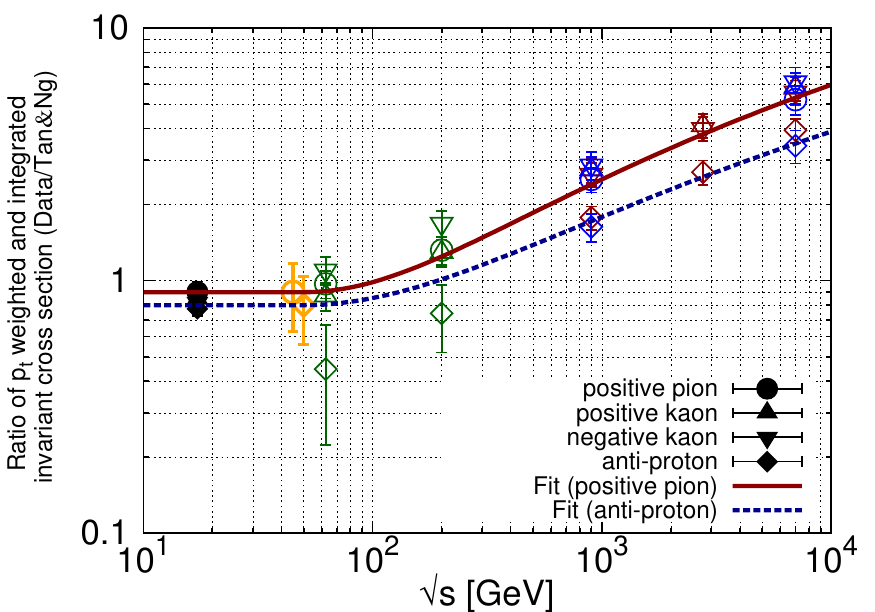}
\caption{
$\sqrt{s}$ dependence of
ratios of $p_t$-weighted cross sections for $\pi^+$, $K^\pm$ and $\bar{p}$
between high energy experiments and Tan\&Ng~\cite{Tan:1984ha}.
Solid lines indicate the correction functions $\xi_H$.
Black, orange, green, blue and red points correspond to NA49, PHENIX, ALICE and CMS data respectively, with estimated systematic uncertainties.
The yellow points represent data sets used in Tan\&Ng fitting paper~\cite{Tan:1984ha}.
}\label{fig:zsqs}
\end{figure}

Next, we analyse the high energy data to determine the behaviour of $\xi_H$ at large $\sqrt s$.  
The scaling factors $\xi_H$ are calibrated to reproduce the $p_t$-weighted cross section of Eq.~(\ref{eq:avp}) evaluated on the high energy experimental data. 
Fig.~\ref{fig:zsqs} shows the 
$\sqrt{s}$ dependence of
ratios of $p_t$-weighted cross sections for $\pi^+$, $K^\pm$ and $\bar{p}$
between high energy data and the Tan\&Ng~\cite{Tan:1984ha} formulae. 
Solid lines indicate the correction functions Eqs.~(\ref{eq:xi_pi}-\ref{eq:xi_K}).
We use data from PHENIX~\cite{Adare:2011vy} at $\sqrt{s}=62.4,~200$ GeV,
CMS~\cite{Chatrchyan:2012qb}  at $\sqrt{s}=900,~2760,~7000$ GeV and
ALICE~\cite{Aamodt:2011zj,Adam:2015qaa} at $\sqrt{s}=900,~7000$ GeV. 

We calculate the $p_t$-weighted cross section using the $p_t$ range provided by the experiments. 
Since CMS and ALICE gives a production yield,
we use a fitting function of inelastic total scattering rate
in~\cite{Winkler:2017xor} to obtain the cross section.
In addition, to obtain inelastic yield for CMS, we multiply an empirical factor $0.78$
(see~\cite{Chatrchyan:2012qb}).
For ALICE data, we use $dN/dy$ estimated in~\cite{Aamodt:2011zj,Adam:2015qaa}.
Statistical and systematic errors are roughly 10\% for all experiments
apart from the PHENIX $\pb$ data, to which we refer in more detail below.

The orange points in Fig.~\ref{fig:zsqs}  summarise the collection of experimental data  used by the Tan\&Ng original analysis~\cite{Tan:1984ha}.
These early measurements cover a wide range of phase space and energy, corresponding to 
$\sqrt{s}\simeq 10-60$ GeV. Detailed comparison shows that the Tan\&Ng fits are consistent with these data to 
within $\sim \pm30\%$, comparable to the internal variation between the results of individual analyses in this data set, and we assign this uncertainty to the orange points.

We find that the correction functions 
\begin{align}
         \xi_{\pi^+}(\sqrt{s})&=\left\{\begin{array}{ll}
0.9 & (\sqrt{s} < 50~{\rm GeV}) \\
0.9 + 0.18 [\log(\sqrt{s} / 50~{\rm GeV})]^2  & (\sqrt{s} \geq 50~{\rm GeV})\end{array}\right., \label{eq:xi_pi}\\
 \xi_{\bar{p}}(\sqrt{s})&=\left\{\begin{array}{ll}
0.8 & (\sqrt{s} < 50~{\rm GeV}) \\
0.8 + 0.11 [\log(\sqrt{s} / 50~{\rm GeV})]^2  & (\sqrt{s} \geq 50~{\rm GeV})\end{array}\right., \label{eq:xi_pbar}\\
       \xi_{K^{\pm}}(\sqrt{s})&=\xi_{\pi^+}(\sqrt{s}), \label{eq:xi_K}
\end{align}
reproduce the experimentally determined $p_t$-weighted cross sections in the range $\sqrt{s} \leq 7$ TeV.
%%%%%%%%%%%%%%%%%%%%%%%%%%%%%%%%%%%%%%%%

Several comments are in order.
First, the PHENIX $\pb$ data~\cite{Adare:2011vy} in Fig.~\ref{fig:zsqs} exhibit larger uncertainty compared to most of the other measurements, and the central values 
are indeed correspondingly off by $\sim 50\%,~30\%$ for $\sqrt{s}$=62.4 and 200 GeV from the fit.
To estimate the $p_t$-weighted $\pb$ cross section from~\cite{Adare:2011vy} we start with the data without feed-down correction, 
as the feed-down corrected cross section is found to be lower by a factor of a few  in
low $p_t$ bins, which appears broadly inconsistent with the remaining data set. 
To estimate the feed-down corrected result, we subtract $30\%$ off the inclusive result, as suggested by our analysis in App.~\ref{app:ant}. 
The $\pb$ systematic uncertainties quoted in~\cite{Adare:2011vy}
 are sizeable, notably in the lower $p_t$ region, 
due to the feed-down correction
 and take maximally 
$\sim 50\%,~30\%$ for $\sqrt{s}$=62.4 and 200 GeV.
In Fig.~\ref{fig:zsqs}, we assign these conservative uncertainty estimates of $50\%$ and $30\%$ to these data. 
 In addition to the feed-down uncertainty, the $p_t$ range covered by the $\pb$ cross section data in~\cite{Adare:2011vy}
 is limited, starting from $p_t=0.6$ GeV. This means that the $p_t$-weighted cross section estimate derived from these data is based on a kinematically sub-dominant region for astrophysical purposes. 

Second, we comment on the $K_S$ contribution to the $\pi$ cross section.
In the analysis of Fig.~\ref{fig:zsqs}, we assume that the $\pi$ cross sections reported by the experiments are 
prompt and do not include $\pi$ from $K_S$ decay.
The NA49 and CMS experiments explicitly state that $\pi$ from $K_S$ decay are discriminated in their analyses.
On the other hand, the treatment in the PHENIX and ALICE experiment is unclear.
This makes 5 \% ambiguity of the points from PHENIX and ALICE experiments in Fig.~\ref{fig:zsqs}.
In practice, this ambiguity is not quantitatively important for the determination of fitting formula.

Finally, we comment on the $x_R$ dependence in the high $\sqrt{s}$ regime.
The high energy experimental data from~\cite{Adare:2011vy,Chatrchyan:2012qb,Aamodt:2011zj,Adam:2015qaa} is only specified at mid-rapidity $(x_R\simeq 0)$.
This means that our fit could fail to reproduce the $x_R$ dependence in the high $\sqrt{s}$ regime. 
Fixing this caveat would require cross section data at non-zero $x_R$ (forward region) in the high $\sqrt{s}$ regime.

\subsection{Comparison to previous work}
\begin{figure}
\centering
\includegraphics[width=\hsize]{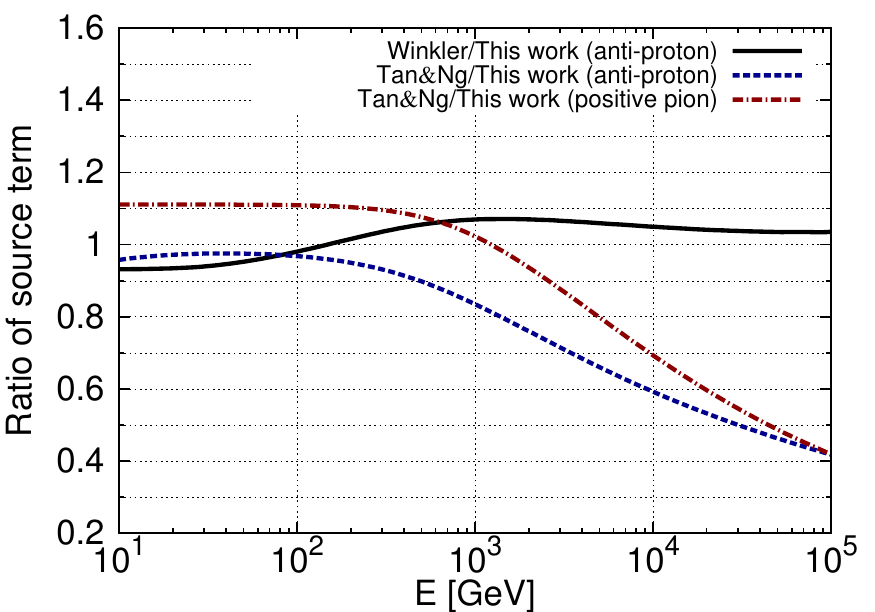}
\caption{
$\sqrt{s}$ dependence of
The ratios of source term assuming $J_p\propto E_p^{-3}$.
The Black line shows the $\pb$ source term ratio between that obtained using the fit of Ref.~\cite{Winkler:2017xor} (denoted ``Winkler'')
and ours. 
The blue dotted (red dashed) line shows the $\bar{p}$ ($\pi^+$)  source term ratio  between Tan\&Ng~\cite{Tan:1984ha} and ours.}\label{fig:zflux}
\end{figure}

In Fig.~\ref{fig:zflux} we show the secondary source terms for $\bar p$ and $\pi^+$, assuming $pp$ production from a power-law primary proton flux $J_p\propto E_p^{-3}$, comparing our results to the fitting formulae of~\cite{Winkler:2017xor} and Tan\&Ng. 
For $\bar p$ production, we now include the contributions from both hyperon decay and decay in flight of $\bar n$,
 using the procedure defined in App.~\ref{app:ant}.  %\cite{Winkler:2017xor}. 
The Black line shows the $\pb$ source term ratio between that obtained using the fit of Ref.~\cite{Winkler:2017xor} (denoted ``Winkler'')
and ours. We find agreement to the 10\% level. 
The blue dotted (red dashed) line shows the $\bar{p}$ ($\pi^+$)  source term ratio  between Tan\&Ng~\cite{Tan:1984ha} and ours. 
The deviation from radial scaling, assumed in Tan\&Ng, is clear at high energy.

\section{The $\ep/\pb$ flux ratio}\label{sec:general_bound}
Ref.~\cite{Katz:2009yd} pointed out that the production rate ratio $Q_{\ep}/Q_{\pb}$ provides a model-independent  upper bound to the flux ratio of high-energy secondary CR $e^+$ and $\bar p$:
\begin{align}
\frac{J_{e^+}({\cal R})}{J_{\bar p}({\cal R})}
< \frac{Q_{e^+}({\cal R})}{Q_{\bar p}({\cal R})},
\end{align}
where the source terms for secondary $\bar p$ and $e^+$ produced in $pp$ collisions are\footnote{Note that (i) the factor 2 in $Q_{\bar p}$ comes from decay in flight of $\bar n$, and (ii) the normalization in our definition for $Q_{\ep,\pb}$ here is somewhat different than in, e.g., Refs.~\cite{Katz:2009yd,Blum:2013zsa,Blum:2017qnn}. This is for ease of presentation and is of no consequence for the source ratio.}:
\begin{align}
Q_{\bar p}(E_{\bar p}) &= 2\int^\infty_{E_{\bar p}} dE_p 4\pi J_p(E_p) \frac{d\sigma_{pp\to \bar p}}{dE_{\bar p}}(E_p; E_{\bar p}), \\
Q_{e^+}(E_{e^+}) &= \int^\infty_{E_{e^+}} dE_p 4\pi J_p(E_p) \frac{d\sigma_{pp\to \bar e^+}}{dE_{e^+}}(E_p; E_{e^+}).
%Q_{\bar p}({\cal R}) &= 2\int d{\cal R}_p 4\pi J_p({\cal R}_p) \frac{d\sigma_{pp\to \bar p}}{dp_{\bar p}} (p_p = {\cal R}_p, p_{\bar p} = {\cal R} ), \\
%Q_{e^+}({\cal R}) &= \int d{\cal R}_p 4\pi J_p({\cal R}_p) \frac{d\sigma_{pp\to \bar e^+}}{dp_{e^+}}  (p_p = {\cal R}_p, p_{e^+} = {\cal R} ),
\end{align}
This upper bound only depends on the inclusive production cross sections and the shape of proton cosmic ray flux $J_p$.

\begin{figure}
\centering
\includegraphics[width=\hsize]{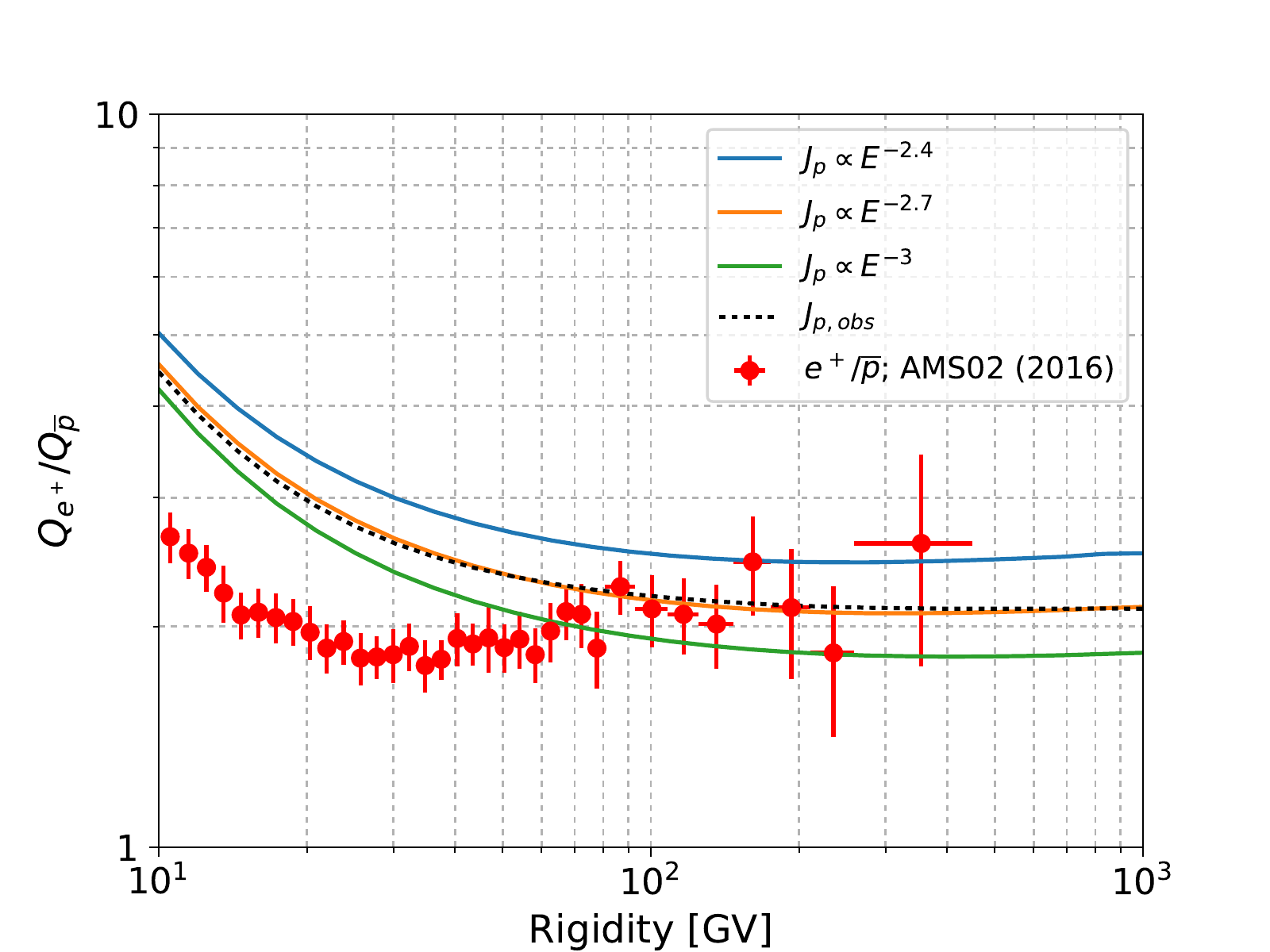}
\caption{The ratio of $Q_{e^+} / Q_{\bar p}$.
The dashed line is calculated by using the observed proton flux.
Blue, orange, and green solid lines are calculated by assuming $J_p \propto E^{-2.4},~E^{-2.7},E^{-3}$, respectively.
The observational data $e^+ / \bar p$ is taken from Ref.~\cite{Aguilar:2016kjl}.
}\label{fig:ratio_of_sourceterm}
\end{figure}

We are now in position to extend the calculation of $Q_{e^+}(\R)/Q_{\bar p}(\R)$ to high energy, and compare with the latest CR data. 
In Fig.~\ref{fig:ratio_of_sourceterm} we show the upper bound predicted for different assumptions on the primary proton flux in the spallation region. The $e^+/\bar p$ flux ratio measured by AMS-02 is consistent with the upper bound and saturates it at high energy (for proton flux coinciding with the locally measured proton flux).

Recent calculations of the high-energy secondary CR $\bar p$ flux~\cite{Blum:2013zsa,Blum:2017qnn}, using up to date $\bar p$ production cross section consistent with our results here and calibrated to agree with AMS-02 B/C data, are consistent with the CR $\bar p$ flux measured by AMS-02. These results are reproduced in App.~\ref{app:secpb}. The significance, in connection with Fig.~\ref{fig:ratio_of_sourceterm}, is that the observed flux of CR $e^+$ at $\R>100$~GV coincides with the expected flux of secondary $e^+$, that would be expected if radiative energy loss became unimportant in the propagation at these energies. Achieving such low level of energy loss would require that the characteristic secondary CR propagation time drops below a few Myr at $\R>100$~GV. 

A comparison of the source ratio $Q_{e^+} / Q_{\bar p}$ to the observed $\ep/\pb$ flux ratio was also presented in Ref.~\cite{Lipari:2016vqk},
which found results for $Q_{e^+} / Q_{\bar p}$ smaller than our value by $\sim 30 \%$ in the energy range 10 -- 1000 GeV. This led Ref.~\cite{Lipari:2016vqk} to argue that $\ep$ energy losses may be negligible at all energies (rather than only at $E\gtrsim100$~GeV, as suggested by our Fig.~\ref{fig:ratio_of_sourceterm}). We have not been able to reproduce the origin of this discrepancy. 
%They used $\bar p$ cross section by Tan\&Ng \cite{Tan:1982nc}, $\pi$ cross section by Badhwar \textit{et.~al.~}\cite{Badhwar:1977zf}, and $K$ cross section by NA49 \cite{Anticic:2010yg}.

\section{Conclusions}\label{sec:conclusion}
We presented an analysis of inclusive $\pb$, $\pi$, and $K$ production in pp collisions. Our main goal was to implement recent experimental data for meson production, in particular the effect of radial scaling violation manifest at LHC energies and recent detailed kinematical data from the NA49 experiment at intermediate energy, in semi-analytic fits used for the calculation of the astrophysical secondary production of $\ep$. We provide fitting formulae that, combined with earlier results from Tan\&Ng~\cite{Tan:1984ha}, allow to compute the astrophysical production of $\ep$ and $\pb$ up to the multi-TeV range with an estimated uncertainty of $\sim20$\%.

The $\ep/\pb$ flux ratio reported by AMS-02 is found to coincide with  the secondary source production rate ratio $Q_{\ep}/Q_{\pb}$ at high-energy $E>100$~GeV. This coincidence may be considered as a hint for a secondary origin for CR $\ep$ and $\pb$, as it would be a fine-tuned accident in models that advocate new primary sources for either antimatter CR species. 

%We discussed two effects on inclusive $e^+$ production cross section which is relevant to cosmic ray study;
%(1) the violation of radial scaling at high energy,
%(2) contributions from cascade decay of $K_L^0$.
%Each of them affects 5--10 \% in $e^+$ cosmic ray flux.
%We also calculated a generic bound on the ratio between $J_{e^+}$ and $J_{\bar p}$.

%\vspace{1mm}~\\

%\vspace{0.1mm}
%~\\
\section*{Acknowledgements}
This
research is supported by grant 1937/12 from 
the I-CORE program of the Planning and Budgeting Committee and the Israel
Science Foundation and by grant 1507/16
from the Israel Science Foundation. The work
of MT is supported by the JSPS Research Fellowship
for Young Scientists. KB is incumbent
of the Dewey David Stone and Harry Levine career
development chair.

\appendix
\section{Neutral kaon contributions}\label{sec:kaon}
In this section we calculate the final state $\ep$ contribution coming from the decay of $K_L^0$ mesons.
This contribution has been neglected in the literature, although the corresponding cross section is comparable to that for charged kaons which was previously taken into account.
$K_L^0$ mesons are long-lived ($c\tau_{K_L^0} \simeq 15~{\rm m}$) in the collider set-up,
so that $\pi^+$ from $K_L^0$ decay are not included in the fitting formula of the inclusive $\pi^+$ cross section.
In addition, the $K_L^0$ semi-leptonic decay contributes directly to $e^+$ and $\mu^+ (\to e^+)$ production.

We consider the following decay channels \cite{Olive:2016xmw}:
\begin{align}
{\rm Br}(K_L^0 \to \pi^\pm e^\mp \nu_e) &= 40.55~\%,\\
{\rm Br}(K_L^0 \to \pi^\pm \mu^\mp \nu_e) &= 27.04~\%,\\
{\rm Br}(K_L^0 \to \pi^+ \pi^- \pi^0) &= 12.54~\%.
\end{align}
We approximate and simplify the kinematics of $K_L^0$ three-body decays, assigning each of the decay products an energy of $m_K/3$ in the $K_L^0$ rest frame and ignoring muon polarisation.
We approximate the $K_L^0$ production cross section to match that of $K^+$.
The $e^+$ spectrum from boosted $\mu^+$ is given in Ref.~\cite{scanlon1965energy} and the $e^+$ spectrum from boosted $\pi^+$ is given in Ref.~\cite{dermer1986binary}. 

\begin{figure}
\centering
\includegraphics[width=\hsize]{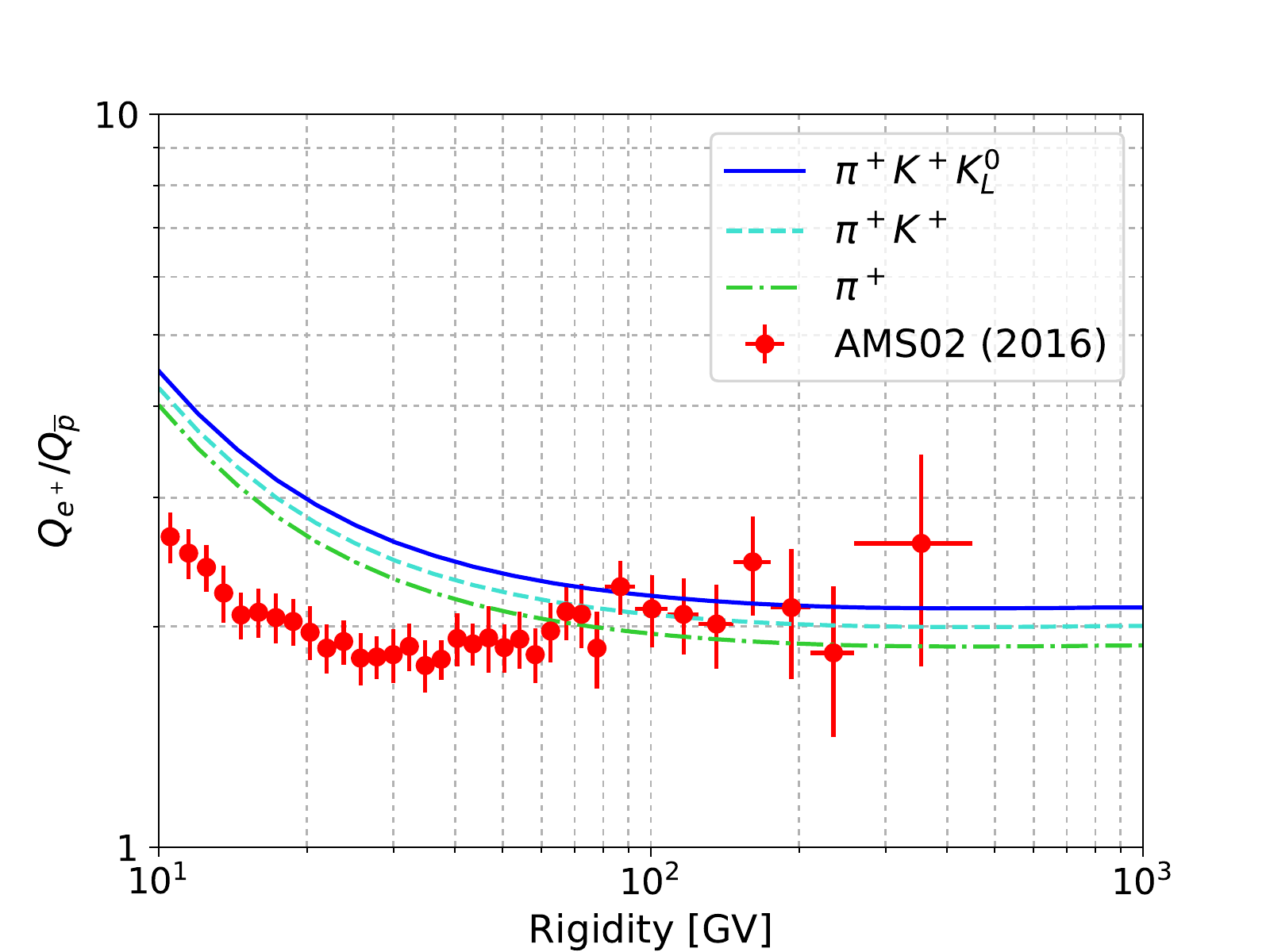}
\caption{Same as in Fig.~\ref{fig:ratio_of_sourceterm}, but highlighting the kaon contribution to the source term of secondary $\ep$. The proton spectrum is assume to follow $J_p\propto E^{-2.7}$.
}\label{fig:positron}
\end{figure}

The kaon contribution to astrophysical secondary $\ep$ production is highlighted in Fig.~\ref{fig:positron}. The 
$K_L^0$ contribution amounts roughly to 5\% of the total $\ep$ source.

\section{Antiproton cross section including anti-hyperon contributions}
\label{app:ant}
In this section we analyse the hyperon contribution to the inclusive $\pb$ production cross section.
We denote the Lorentz-invariant differential cross section as $f$;
\begin{align}
         f^\bullet_\# \equiv E\frac{d^3\sigma^\bullet_\# }{dp^3}.
\end{align}
The astrophysically relevant inclusive 
$f_{\bar{p}}^{\rm tot}$, which includes
effects from $\bar n$ and hyperon decays, can be decomposed in the following way:
\begin{align}
         f_{\bar{p}}^{\rm tot}&=f_{\bar{p}}+f_{\bar{n}},\\
         f_{\bar{p}}&=f_{\bar{p}}^0+f_{\bar{p}}^{\bar{\Lambda}}+f_{\bar{p}}^{\bar{\Sigma}},\\
         f_{\bar{n}}&=f_{\bar{n}}^0+f_{\bar{n}}^{\bar{\Lambda}}+f_{\bar{n}}^{\bar{\Sigma}},
\end{align}
where $f^0$ indicates the prompt contribution
and $f^{\bar{\Lambda}, \bar\Sigma}$ denote contribution from the hyperon decay.
Neglecting isospin violation, we assume $f_{\bar{p}}^0=f_{\bar{n}}^0$.
%As a reference cross-section, we take Tan\&Ng antiproton cross section $f_{\bar{p}}^{\rm Tan}$.
To set a rough scale for the effect we're after here, the analysis in Sec.~\ref{sec:radialscaling} shows $f^0_{\bar{p}}\simeq 0.8f^{\rm Tan\&Ng}_{\bar p}$ at low $\sqrt{s}$, where $f_{\bar{p}}^{\rm Tan\&Ng}$ includes the hyperon decay contribution.
(See Eq.~(\ref{eq:xi_pbar}) and Fig.~\ref{fig:zsqs}.)

\subsection{Anti-hyperon production cross section at the NA49 experiment}
NA49 \cite{Alt:2005zq} results indicate that the kinematical distribution of anti-hyperons produced in pp collisions is somewhat different from that of anti-nucleons\footnote{This conclusion is in some tension with the parallel discussion presented in~\cite{Winkler:2017xor}, which will lead us to slightly different results.}.
We introduce $x_R$-dependent functions $g_{\bar B}(x_R)$ with $\bar B = \bar \Lambda, \bar\Sigma^\pm$,
and parametrize the hyperon contributions as
\begin{align}
         f_{\bar{p}}^{\bar{\Lambda}} =& f^{\rm Tan\&Ng}_{\bar{p}} g_{\bar \Lambda}(x_R)
                                         \text{Br} ({\Lambda} \rightarrow {p}X),\\
         f_{\bar{n}}^{\bar{\Lambda}} =& f^{\rm Tan\&Ng}_{\bar{p}} g_{\bar \Lambda}(x_R)
                                         \text{Br} ({\Lambda} \rightarrow {n}X),\\
         f_{\bar{p}}^{\bar{\Sigma}}  =& f^{\rm Tan\&Ng}_{\bar{p}} g_{\bar \Sigma^-}(x_R)
                                         \text{Br} ({\Sigma}^+ \rightarrow {p}X),\\
         f_{\bar{n}}^{\bar{\Sigma}}  =& f^{\rm Tan\&Ng}_{\bar{p}} g_{\bar \Sigma^-}(x_R)
                                         \text{Br} ({\Sigma}^+ \rightarrow {n}X) \nonumber\\
                                      &+f^{\rm Tan\&Ng}_{\bar{p}} g_{\bar \Sigma^+}(x_R)
                                         \text{Br}({\Sigma}^- \rightarrow {n}X).
\end{align}
The branching fractions for hyperon decays are 
$\text{Br}({\Lambda} \rightarrow {p}X)\simeq 0.64$,
$\text{Br}({\Lambda} \rightarrow {n}X)\simeq 0.36$, $\text{Br}({\Sigma}^+ \rightarrow {p}X)\simeq 0.52$,
$\text{Br}({\Sigma}^+ \rightarrow {n}X)\simeq 0.48$ and
$\text{Br}({\Sigma}^- \rightarrow {n}X)\simeq 1$ \cite{Patrignani:2016xqp}.
Summing up, we obtain
\begin{align}
         %f^{\rm tot}_{\bar{p}}\simeq f^{\rm Tan}_{\bar{p}}
         %\times\left[1.6+\frac{\bar{\Lambda}}{\bar{p}_{\rm Tan}}+
         %\frac{\bar{\Sigma}^-}{\bar{p}_{\rm Tan}}+
         %\frac{\bar{\Sigma}^+}{\bar{p}_{\rm Tan}}
         %\right].
         f^{\rm tot}_{\bar{p}}\simeq f^{\rm Tan}_{\bar{p}}
         \times\left[1.6+
         g_{\bar{\Lambda}}+
         g_{\bar{\Sigma}^-}+
         g_{\bar{\Sigma}^+}
         \right]. \label{eq:correction at NA49}
\end{align}
We neglect momentum difference between parent and daughter particle
since their mass difference is $\lesssim$20\%.

Let us determine $g_{\bar B}(x_R)$.
NA49 analysis~\cite{Alt:2005zq} (see Fig. 22 there)
offers the differential multiplicity $dn/dx_F$ for $\Lambda,~\bar{\Lambda},~\Sigma^+,~\Sigma^-$, defined as
\begin{align}
         \frac{dn_\bullet}{dx_F}(x_F)=\frac{\pi}{\sigma^{\rm inel}}\frac{\sqrt{s}}{2}
         \int dp_t^2 \frac{f_\bullet}{E}.
\end{align}
Uncertainties of $dn/dx_F$ are not presented, but a typical error estimate of $\sim 20\%$
can be inferred from the analysis in~\cite{Winkler:2017xor}.

Although the definition of $x_R (= E^* / E_{\rm max}^*)$ and $x_F (= 2 p_L^* / \sqrt{s})$ are different,
their difference is of the order of $p_t^2/s$ or $m_p^2/s$.
Thus, $g_{\bar B}(x_R)$ can be determined from the observation of $dn/dx_F$.
As discussed in section \ref{sec:radialscaling}, $0.1 \lesssim x_R \lesssim 0.4$ is the important kinematical region to determine secondary cosmic ray production.
In this region, the $p_t$ dependence on $E$ becomes weak and $dn/dx_F$ is determined by $p_t$ weighted averaged cross-section.
In this respect, we find that $dn/dx_F$ is a directly relevant quantity for secondary cosmic ray production. 
Then, it is reasonable to estimate 
\begin{align}
         g_{\bar B}(x_R) = \left[ \left( \frac{dn_{\bar B}}{dx_F} \right) \bigg/ \left( \frac{dn_{\bar p}}{dx_F} \bigg|_{\rm Tan\&Ng} \right) \right] \Bigg|_{x_F=x_R},
\end{align}
with $\bar B=\Lambda,~\bar{\Lambda},~\Sigma^+,~\Sigma^-$ or $\bar{p}$.

Following Ref.~\cite{Anticic:2009wd} we assume the relation
\begin{align}
         \frac{dn_{\bar{\Sigma}^-}}{dx_F} \simeq 0.8\times \frac{dn_{\bar\Lambda}/dx_F}{dn_\Lambda/dx_F} \frac{dn_{\Sigma^+}}{dx_F}.
\end{align}
Then, we expect
\begin{align}
         g_{{\bar{\Sigma}^-}} \simeq 0.8\times \frac{dn_{\bar\Lambda}/dx_F}{dn_\Lambda/dx_F} g_{\Sigma^+}.
\end{align}
We assume a similar relation for $\bar{\Sigma}^+$:
\begin{align}
         g_{{\bar{\Sigma}^+}} \simeq 0.8\times \frac{dn_{\bar\Lambda}/dx_F}{dn_\Lambda/dx_F} g_{\Sigma^-}.
\end{align}

To obtain $g_{\bar B}$ (with $B=\Lambda,~\Sigma^{\pm})$,
we fit the $x_F$ dependence shown in the NA49 analysis by the following form:
\begin{align}
         g_{\bar B}=a(1-x_R)^n.
\end{align}
We found $(a,n)$=$(0.13,-3)$, $(0.038,-3)$, $(0.028,-2)$ well fit $\bar{\Lambda}$,
$\bar{\Sigma}^-$, $\bar{\Sigma}^+$ respectively.

\begin{figure}
\centering
\includegraphics[width=\hsize]{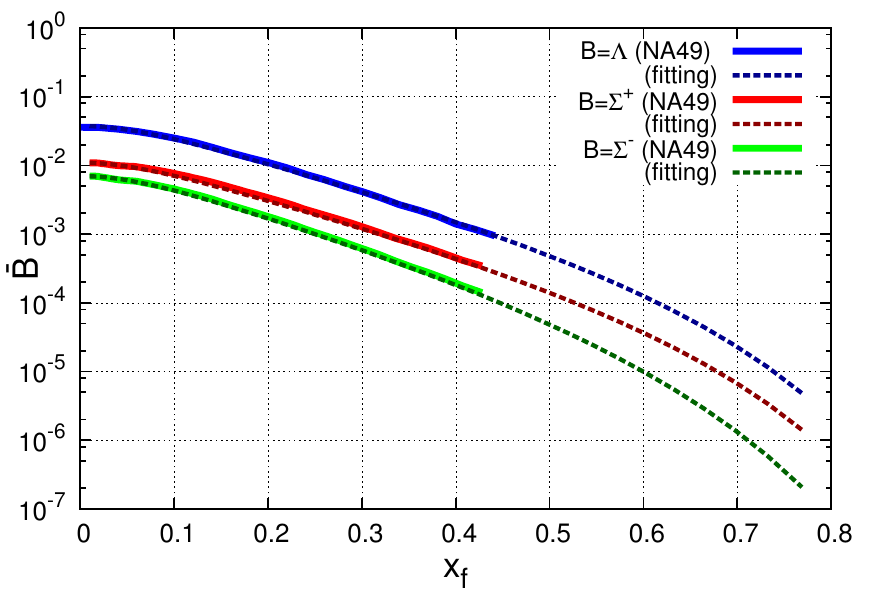}
\caption{
$x_F$ dependent $\bar{B}\equiv dn_{\bar{B}}/dx_F$. 
The solid lines 
and dashed ones correspond to NA49 values and our fitting function, respectively.
}\label{fig:zhyp}
\end{figure}

Fig.~\ref{fig:zhyp} shows the $x_F$ dependent $\bar{B}\equiv dn_{\bar{B}}/dx_F$. 
Solid and dashed lines correspond to NA49 values and our fitting function, respectively.

%\textcolor{red}{\bf *MT: we should comment on the error$\sim 20\%$.}\\

%%%%%%%%%%%%%%%%%%%%%%%%%%%%%%%%%%%%%%%%%%%%%%%%%
\begin{figure}
\centering
\includegraphics[width=\hsize]{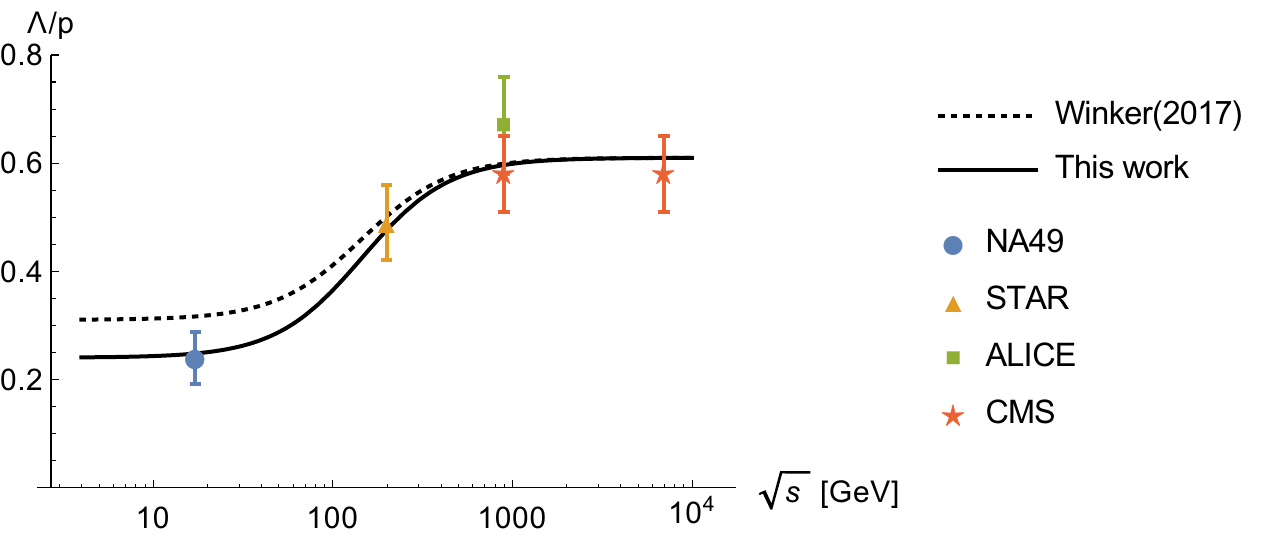}
\caption{
$\bar\Lambda / \bar p$ ratio in proton-proton collision at mid-rapidity.
}\label{fig:lambdapbar}
\end{figure}

\subsection{Multiplicity of anti-hyperons at large $\sqrt s$}
For relatively small $\sqrt{s}<50$~GeV, we expect that Eq.~(\ref{eq:correction at NA49}) holds with weak $\sqrt s$ dependence.
This is because, empirically, radial scaling applies at small $\sqrt s$.
However, when we consider large $\sqrt s>50$~GeV, we have to consider the violation of radial scaling.

Ref.~\cite{Winkler:2017xor} showed that the ratio between the multiplicity of anti-hyperons and $\bar p$ is not constant as function of $\sqrt{s}$.
Following~\cite{Winkler:2017xor}, we introduce $\sqrt s$ dependence as an overall factor to the hyperon contributions,
\begin{align}
         %f^{\rm tot}_{\bar{p}}\simeq f^{\rm Tan}_{\bar{p}}
         %\times\left[1.6\times \xi(\sqrt{s})+
         %\left(\frac{\bar{\Lambda}}{\bar{p}_{\rm Tan}}+
         %\frac{\bar{\Sigma}^-}{\bar{p}_{\rm Tan}}+
         %\frac{\bar{\Sigma}^+}{\bar{p}_{\rm Tan}}\right)\times \kappa(\sqrt{s})
         %\right],
         f^{\rm tot}_{\bar{p}}\simeq f^{\rm Tan}_{\bar{p}}
         \times\left[1.6\times \xi(\sqrt{s})+
         \left(g_{\bar{\Lambda}}+
         g_{\bar{\Sigma}^-}+
         g_{\bar{\Sigma}^+}\right)\times \kappa(\sqrt{s})
         \right].
\end{align}
Here $\kappa(\sqrt s)$ satisfies $\lim_{s\to 0} \kappa(\sqrt s) = 1$, and deviates from unity at large $\sqrt s$.

We define the ratio between the multiplicity of $\bar\Lambda$ and $\bar p$ at midrapidity:
\begin{align}
\frac{\bar\Lambda}{\bar p} = \frac{ dn_{\bar \Lambda}/dx_F }{dn_{\bar p}/dx_F} \bigg|_{x_F=0}.
\end{align}
For simplicity, we assume that $\bar\Lambda,~\bar\Sigma^\pm$ have the same scaling law for their multiplicity.
By using this assumption, we take $\kappa(\sqrt s)$ as
\begin{align}
\kappa(\sqrt s) = \frac{(\bar\Lambda/\bar p)(\sqrt s)}{(\bar\Lambda/\bar p)(0)}.
\end{align}

Finally, we analyse the ratio $\bar\Lambda/\bar p$ using data from  
STAR \cite{Abelev:2006cs, Abelev:2008ab},
ALICE \cite{Aamodt:2011zza, Aamodt:2011zj},
and CMS \cite{Khachatryan:2011tm, Chatrchyan:2012qb} which 
provided multiplicity ratios at mid-rapidity.
NA49 also provided differential multiplicity at the mid-rapidity;
%\begin{align}
%dn_{\bar p}/dx_F|_{x_F=0} &= 0.1477\qquad  \cite{Anticic:2009wd}.\\
%dn_{\bar\Lambda}/dx_F|_{x_F=0} &= 0.036\qquad  \cite{Alt:2005zq}.
%\end{align}
we assume an uncertainty of 20\% from the uncertainty in the feed-down correction. 
This gives us $\bar\Lambda / \bar p = 0.24 \pm 0.05$ at $\sqrt{s} = 17.2~{\rm GeV}$ form NA49 experiment.

Fig.~\ref{fig:lambdapbar} shows our result.
For comparison, we also show $\bar\Lambda/\bar p$ as found in~\cite{Winkler:2017xor}.
Our results can be fitted by the following formula:
\begin{align}
\frac{\bar\Lambda}{\bar p} = 0.24 + \frac{0.37}{1 + ((146~{\rm GeV})^2/s)^{0.9}}.
\end{align}

%%%%%%%%%%%%%%%%
\section{Secondary $\pb$}\label{app:secpb}
\begin{figure}
\centering
\includegraphics[width=\hsize]{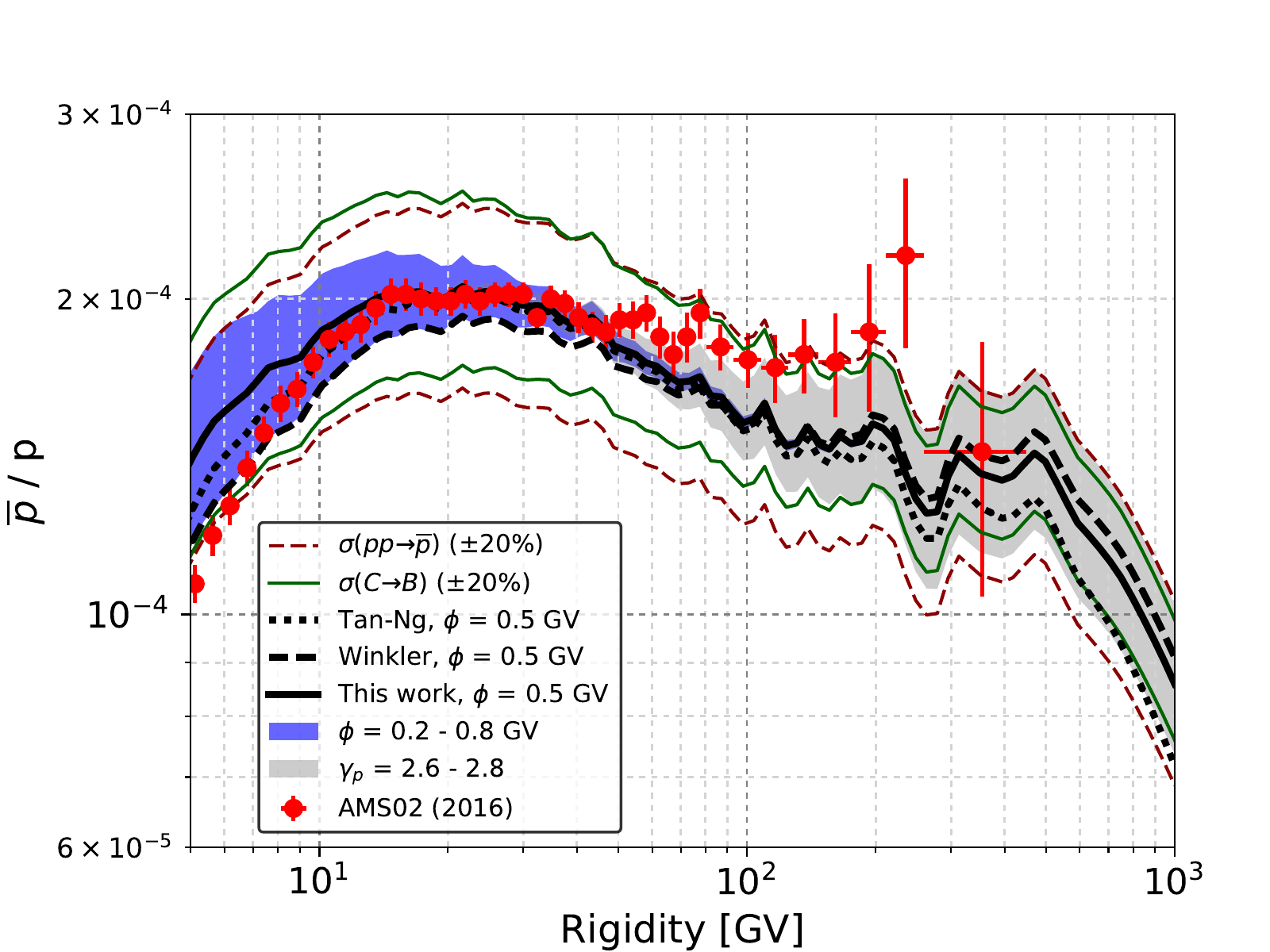}
\caption{
Cosmic ray $\bar p$ flux for several cross section formulae.
In all cases, we use the mean traversed target column density extracted from B/C data~\cite{Katz:2009yd} using nuclear fragmentation cross sections as specified in~\cite{Blum:2017qnn}.
Solid black line shows the prediction using our fit.
Dotted and dashed black lines show to the result when using the fit from Tan\&Ng \cite{Tan:1984ha} and Winkler \cite{Winkler:2017xor}, respectively.
Other bands and lines show various sources of systematic uncertainty; see text for details.
} \label{fig:antiproton}
\end{figure}

Fig.~\ref{fig:antiproton} shows the secondary $\bar p$ cosmic ray flux predicted by our cross section formula, calculated under the assumption that the mean target column density traversed by CR protons; He; nuclei such as B, C, and O; and $\pb$ is the same as function of magnetic rigidity~\cite{Katz:2009yd}. The column density used in the calculation is extracted from B/C data using fragmentation cross sections as specified in~\cite{Blum:2017qnn}. 

The simple estimate in Fig.~\ref{fig:antiproton} is consistent the AMS-02 $\bar p$ data~\cite{Aguilar:2016kjl}. The calculation is sensitive to a number of systematic uncertainties.
The blue region shows the uncertainty of the solar modulation parameter $\phi = (0.2-0.8)$ GV.
The grey region shows the result of varying the spectral index of proton CR above 300 GV.
We vary $\gamma_p$ in the range of 2.6--2.8 where $J_p \propto E_p^{-\gamma_p}$: this should represent the possibility that the CR proton spectrum in the regions dominating secondary $\pb$ production may not be identical to the locally measured spectrum.
The solid green lines show the result of varying the C$\to$B fragmentation cross section by $\pm 20\%$. 
Finally, the dashed dark lines represent $\bar p$ production cross section uncertainty of $\pm 20\%$.

\bibliography{ref}
\bibliographystyle{utphys}

\end{document}